\documentclass[twocolumn]{aastex63}
\usepackage{apjfonts}

\received{}
\revised{}
\accepted{}

\submitjournal{ApJ}

\shorttitle{SUB-PARSEC-SCALE FREE-FREE ABSORPTION DISK IN 3C~84}
\shortauthors{WAJIMA, KINO, \& KAWAKATU}

\begin{document}

%%%%%%%%%%%%%
%%% Title %%%
%%%%%%%%%%%%%
\title{Constraints on Circumnuclear Disk through Free-Free Absorption in the
Nucleus of 3C~84 \\ with KaVA and KVN at 43 and 86~GHz}

%%%%%%%%%%%%%%%%%%%%%%%%%%%%
%%% Author & Affiliation %%%
%%%%%%%%%%%%%%%%%%%%%%%%%%%%

%\correspondingauthor{Kiyoaki Wajima}
\email{wajima@kasi.re.kr}

\author{Kiyoaki Wajima}
\affiliation{Korea Astronomy and Space Science Institute,
776 Daedeokdae-ro, Yuseong, Daejeon 34055, Korea}

\author{Motoki Kino}
\affiliation{Kogakuin University of Technology \& Engineering,
Academic Support Center,
2665-1 Nakano, Hachioji, Tokyo 192-0015, Japan}
\affiliation{National Astronomical Observatory of Japan,
2-21-1 Osawa, Mitaka, Tokyo 181-8588, Japan}

\author{Nozomu Kawakatu}
\affiliation{National Institute of Technology, Kure College,
2-2-11 Agaminami, Kure, Hiroshima 737-8506, Japan}

%%%%%%%%%%%%%%%%
%%% Abstract %%%
%%%%%%%%%%%%%%%%
\begin{abstract}

The nearby bright radio galaxy 3C~84 at the center of Perseus cluster is one of
the ideal targets to explore the jet in active galactic nuclei (AGNs) and its
parsec-scale environment.
A recent research of Fujita \& Nagai revealed the existence of the northern
counter-jet component (N1) located at $\sim$ 2~mas north from the central core
in VLBI images at 15 and 43~GHz and they are explained by the free-free
absorption (FFA) due to an ionized plasma foreground.
Here we report a new quasi-simultaneous observation of 3C~84 with the Korean VLBI
Network (KVN) at 86~GHz and the KVN and VERA (VLBI Exploration of Radio
Astrometry) Array (KaVA) at 43~GHz in 2016 February.
We succeeded the first detection of N1 at 86~GHz and the data show that N1 still
has an inverted spectrum between 43 and 86~GHz with its spectral index $\alpha$
($S_{\nu} \propto \nu^{\alpha}$) of $1.19 \pm 0.43$, while the approaching lobe
component has the steep spectrum with the index of $-0.54 \pm 0.30$.
Based on the measured flux asymmetry between the counter and approaching lobes,
we constrain the averaged number density of the FFA foreground $n_{\rm e}$ as
$1.8 \times 10^{4}~{\rm cm^{-3}} \lesssim n_{\rm e} \lesssim 1.0 \times
10^{6}~{\rm cm^{-3}}$.
Those results suggest that the observational properties of the FFA foreground
can be explained by the dense ionized gas in the circumnuclear disk and/or
assembly of clumpy clouds at the central $\sim 1$~pc region of 3C~84.

\end{abstract}

%%%%%%%%%%%%%%%%
%%% Keywords %%%
%%%%%%%%%%%%%%%%
\keywords{galaxies: active --- galaxies: individual (3C~84) ---
radio continuum: galaxies --- techniques: interferometric}

%%%%%%%%%%%%%%%%%%%%%%%%%%%%%%%
%%% Section 1: Introduction %%%
%%%%%%%%%%%%%%%%%%%%%%%%%%%%%%%
\section{Introduction}
\label{sec:Section1}

Active galactic nuclei (AGNs) is widely believed to have an obscuring structure
near a central supermassive black hole (SMBH), which is described in the AGN
unified model \citep[e.g.,][]{Antonucci93,Urry95}.
Indeed, there is mounting evidence that it has a rich structure within 10~pc
scale from the recent progresses of spatially resolved multi-wavelength
observations.
At near- to mid-infrared (NIR to MIR) wavelengths thermal dust emission in AGNs
support the existence of compact obscuring structures within 10~pc of the central
engine \citep[e.g.,][]{Jaffe04,Burtscher13,Asmus14}.
Recent ALMA observations revealed that multi-phase dynamic nature in the
circumnuclear disk (CND) region of larger than 10~pc, i.e., the diffuse atomic
gas is more spatially extended along the vertical direction of the disk than the
dense molecular gas \citep{Izumi18}.
Due to energy feedback from AGNs and nuclear starburst, co-existence of ionized
gas and cold gas is expected on the CND scale \citep[e.g.,][]{Wada16}.
Dense molecular gas disks with their sizes of 10~pc, which may be an outer part
of a few parsec obscuring structure, have been found around nearby Seyfert nuclei
\citep[e.g.,][]{Hicks13,Davies14,Imanishi16,Imanishi18,Izumi18}.
Such a CND would be a massive reservoir of molecular gas, which potentially
triggers an active star formation.
A prominent star formation has been found as a nuclear starburst
\citep[e.g.,][]{Imanishi04,Davies07,Imanishi11,Diamond-Stanic12,
Alonso-Herrero14,Esquej14,Mallmann18}, which may be related to the CND structure
\citep{Kawakatu08,Kawakatu20}.

Despite the above-shown progresses on the scale of larger than a few pc, it is
not yet clear about the physical properties of the obscuring structure on the
scale of smaller than 1~pc.
There are a number of theoretical models arguing possible origins, e.g.,
(1) radiation pressure from AGN \citep[e.g.,][]{Krolik07,Namekata14,Namekata16},
(2) radiation pressure from nuclear starburst \citep[e.g.,][]{Thompson05},
(3) high velocity dispersion clouds/clumps model \citep{Krolik88,Vollmer08},
(4) turbulent pressure from type II supernova explosions \citep{Wada02,Wada09},
(5) disk winds \citep[e.g.,][]{Elitzur06,Nomura16},
(6) outflows driven by AGN radiation pressure
\citep{Wada12,Wada16,Dorodnitsyn16,Chan16,Chan17}, and
(7)  chaotic cold accretion (CCA) within the inner kpc \citep{Gaspari13}.
In order to investigate its physical origin in details on $< 1$~pc scale,
high-resolution observations of the hot ionized gas around AGNs are essential to
understand basic properties of multi-phase CNDs.
As previous researchers have been carried out mainly with very long baseline
interferometer (VLBI) at centimeter wavelengths
\citep[e.g.,][]{Kameno00, Kameno01}, the free-free absorption is one of useful
tools to explore the ionized gas around AGNs.

The compact radio source 3C~84 (also known as NGC~1275) is one of the nearby
($z = 0.018$) best-studied radio galaxies.
Proximity of the object allows us to make detailed observations about the
environment around the SMBH on 1~pc scale.
\citet{Abdo09} reported increase in the radio flux at 14.5~GHz starting in
2005 with long-term monitoring by the the University of Michigan Radio Astronomy
Observatory.
They claimed that this radio flare could be interpreted as an ejection of new jet
components.
This was confirmed by \citet{Nagai10}, who found the emergence of a newborn
bright component, designated as C3, with multi-epoch VLBI monitoring during
2006 -- 2009.
C3 showed a proper motion toward the southern direction with an apparent velocity
of 0.2 -- 0.3$c$ \citep{Nagai10,Suzuki12,Hiura18}.
\citet{Walker00} conducted multi-epoch, multi-frequency VLBA (Very Long Baseline
Array) observations of 3C~84 with the frequency range of 0.3 to 43.2~GHz,
resulting in detection of free-free absorbed emission in the northern lobe
located at $\sim 8$~mas from the core.
They suggested that the absorption feature is due to the existence of 3 pc-scale
absorber.

\citet{Fujita17} firstly reported the existence of the northern counter-jet
component, designated as N1, in 3C~84 at both 15 and 43~GHz with VLBA.
This feature is considered to be a counter jet component corresponding to the
approaching jet located at the south.
N1 has a strongly inverted spectrum, which indicates that it is absorbed by an
ionized plasma around the SMBH via FFA.
So far, however, no detection of N1 has been made at higher frequency than
43~GHz.
Observations at 86~GHz or even higher frequencies offer a unique view on the
environment of the radio jet in 3C~84 since the radio emission would be more
transparent at such frequencies.

In this paper we report the results of VLBI observation of 3C~84 at 86~GHz, which
focuses on properties of newly detected northern component, together with the
quasi-simultaneous 43~GHz image obtained in our previous work summarized in
\citet{Kino18}.
By combining the northern counter lobe images at 86 and 43~GHz, we show the
spectral index between these frequencies and also discuss properties and geometry
of sub-pc-scale structure and the circumnuclear environment in 3C~84.

Throughout this paper, we define the spectral index, $\alpha$, as $S_{\nu}
\propto \nu^{\alpha}$, where $S_{\nu}$ is the flux density at the frequency
$\nu$, and we adopt a $\Lambda$CDM cosmology with
$H_0 = 71$~km~s$^{-1}$~Mpc$^{-1}$, $\Omega_{\Lambda} = 0.73$, and
$\Omega_{\mathrm{M}}= 0.27$ \citep{Komatsu09}, corresponding to an
angular-to-linear scale conversion of 0.36~pc~mas$^{-1}$ for 3C~84.

%%%%%%%%%%%%%%%%%%%%%%%%%%%%%%%%%%%%%%%%%%%%%%%%%%
%%% Section 2: Observations and Data Reduction %%%
%%%%%%%%%%%%%%%%%%%%%%%%%%%%%%%%%%%%%%%%%%%%%%%%%%
\section{Observations and Data Reduction}
\label{sec:Section2}

Our observations were conducted with the KVN (Korean VLBI Network) and VERA (VLBI
Exploration of Radio Astrometry) array (hereafter KaVA) on 2016 February 22 at
43.2~GHz with the total on-source time of 475 minutes, and with KVN on 2016
February 23 at 86.2~GHz with the total on-source time of 737 minutes.
The KaVA 7-telescope data was correlated using the Daejeon Hardware Correlator
\citep{Lee15a} with the output preaveraging time of 1.6 seconds, whereas the DiFX
correlator \citep{Deller11} was used for the KVN 3-telescope data correlation
with the output pre-averaging time of 2 seconds.

The data were reduced using the Astronomical Image Processing System (AIPS)
software \citep{Greisen03} for amplitude and phase calibration, and the Caltech
software Difmap \citep{Shepherd97} for imaging and self-calibration.
We applied {\it a~priori} amplitude calibration using the antenna gain factors
and system noise temperature measurements with the AIPS task APCAL.
We also applied the amplitude correction factor with APCAL, as mentioned by
\citet{Lee15b}.
Bandpass calibration for both amplitude and phase is employed with the AIPS task
BPASS.
Fringe fitting was done using the AIPS task FRING with the solution interval of
30 seconds, resulting in successful fringe detection on all baselines for whole
observing time.
Fringe-fitted data were exported to Difmap for imaging.
We applied the amplitude and phase self-calibration to KaVA's data and could
reconstruct the source model using the visibility for all baselines with the
maximum error of 6\%.
Although we could not apply the amplitude self-calibration to KVN's data
because of small number of antennas, we believe that the visibility amplitude
of KVN's data was calibrated well since dense measurement (every 10 seconds) of
the system noise temperature was made at all KVN stations.
To estimate the amplitude calibration error of KVN's data, we compared the
observation results of a bright quasar 3C~273 obtained by the KVN key science
program, iMOGABA
\citep[interferometric monitoring of gamma-ray bright AGNs;][]{Lee16}, and a VLBA
observation, both of which were conducted quasi-simultaneously (2014 February 28
and 26 for iMOGABA and VLBA, respectively) at 86~GHz by employing the same
amplitude calibration procedure as that of our observation of 3C~84.
As a result, we confirmed that the peak intensity and total CLEANed flux of each
observation are coincident within the range of 15\%.
To ensure a better sensitivity, we adopted natural weighting of the data
with gridding weights scaled by amplitude errors raised to the power of $-1$.
Details of the image dynamic range (DR) obtained with KaVA at 43~GHz have been
reported by \citet{Kino18}, while DR of 100 was obtained for the image at
86~GHz with KVN.

%%%%%%%%%%%%%%%%%%%%%%%%%%
%%% Section 3: Results %%%
%%%%%%%%%%%%%%%%%%%%%%%%%%
\section{Results}
\label{sec:Section3}

\begin{figure*}
\epsscale{1.00}
\plotone{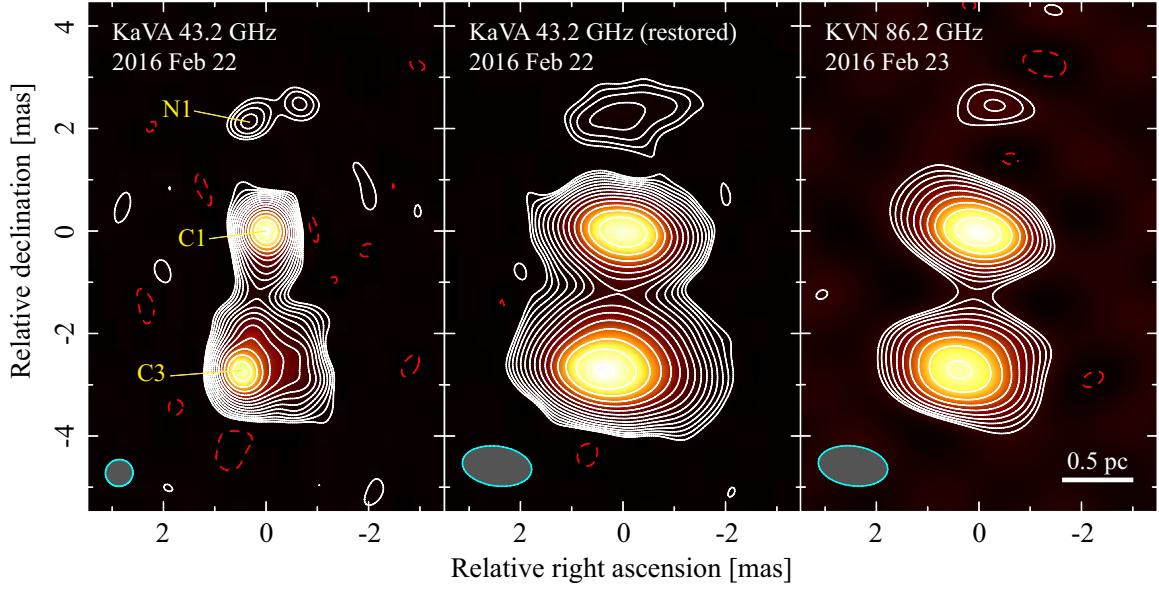}
\caption{VLBI images of 3C~84 obtained by KaVA and KVN.
(Left) KaVA image of 3C~84 on 2016 February 22 at 43.2~GHz.
(Center) Same as the left panel but the synthesized beam is restored with the
same size as that of the right panel.
(Right) KVN image of 3C~84 on 2016 February 23 at 86.2~GHz.
In all figures the lowest contour is three times the off-source rms noise
($\sigma$).
The dashed and solid curves show negative and positive contours, respectively,
and the contour levels are $-3\sigma$, $3\sigma$ $\times$ $(\sqrt2)^n$
($n$ = 0, 1, 2, $\cdot\cdot\cdot$).
The restoring beam with its size of 0.54~mas $\times$ 0.52~mas
at a position angle of $-72^{\circ}$ (left) or 1.36~mas $\times$ 0.78~mas at a
position angle of $82^{\circ}$ (center and right) is indicated in the lower
left corner of each image.}
\label{fig:Figure1}
\end{figure*}

Figure~\ref{fig:Figure1} shows images of 3C~84 with KaVA at 43~GHz and KVN at
86~GHz.
%%%%%%%%%%%%%%%%
%%% Figure 1 %%%
%%%%%%%%%%%%%%%%
Restored KaVA image with the synthesized beam of the KVN 86~GHz image is also
shown in Figure~\ref{fig:Figure1}.
The source consists of two bright components, the central core component
(hereafter C1) and the southern lobe component (hereafter C3), both of which
were identified in the previous VLBI observations
\citep[e.g.,][]{Nagai10,Nagai12,Nagai14}.
On the other hand, we could identify N1 in the north of C1 with the separation
angle of 2.5~mas.
This is already reported by \citet{Fujita17} as mentioned in
Section~\ref{sec:Section1}, whereas we could identify N1 at 86~GHz as well as
43~GHz.
The detection level of N1 is about 6 times the off-source rms noise
($\sigma = 13$~mJy and 65~mJy at 43 and 86~GHz, respectively)
at both frequencies.

\begin{figure*}
\epsscale{0.70}
\plotone{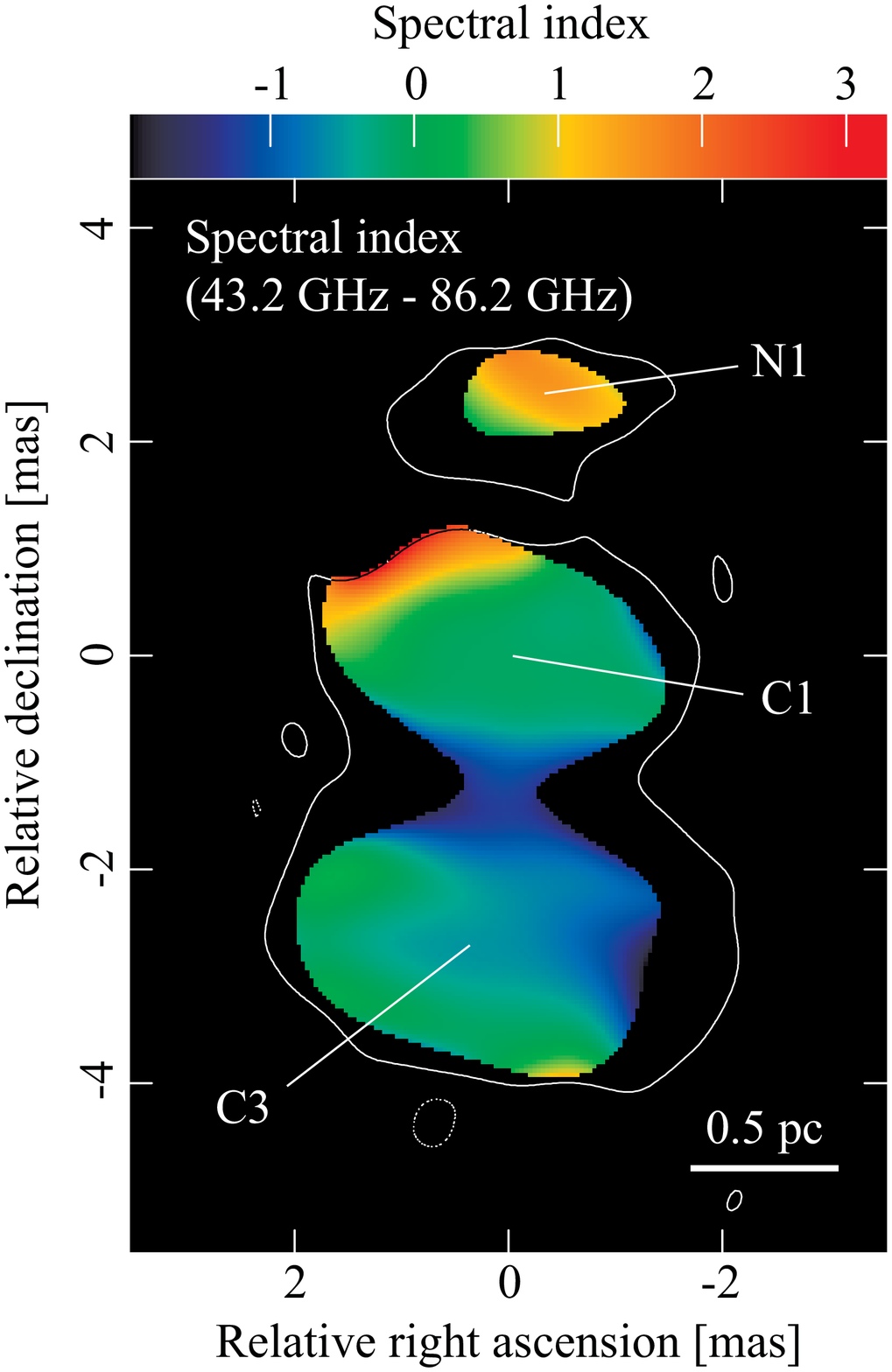}
\caption{Spectral index map of 3C~84 derived from the flux densities at 43.2~GHz
with KaVA on 2016 February 22 and 86.2~GHz with KVN on 2016 February 23.
The top color bar indicates the spectral index $\alpha$ ($S_{\nu} \propto
\nu^{\alpha})$.
The map corresponds to an area greater than $3\sigma$ noise level of the KVN
image at 86.2~GHz, and the solid lines show the lowest contour corresponding to
the $3\sigma$ image noise of the restored KaVA image at 43.2~GHz.}
\label{fig:Figure2}
\end{figure*}

Figure~\ref{fig:Figure2} shows the spectral index map of 3C~84 between KaVA at
43.2~GHz and KVN at 86.2~GHz.
%%%%%%%%%%%%%%%%
%%% Figure 2 %%%
%%%%%%%%%%%%%%%%
Both observations do not employ the phase-referencing technique, resulting in
the loss of the absolute position through the self-calibration procedure
\citep{Pearson84,Thompson01}.
We therefore superposed two images with reference to C3 since it shows an
optically thin feature at the peak intensity position.
To confirm this feature quantitatively, we estimate $\alpha$ of each component.
Table~\ref{tbl:Table1} shows peak intensities of each component at 43.2 and
86.2~GHz, corresponding to the center and right panels in
Figure~\ref{fig:Figure1}, and $\alpha$ between these frequencies.
%%%%%%%%%%%%%%%
%%% Table 1 %%%
%%%%%%%%%%%%%%%
\begin{deluxetable}{cccc}
\tablecolumns{4}
\tabletypesize{}
\tablewidth{0pt}
\tablecaption{Peak intensity and spectral index of each component.\label{tbl:Table1}}
\tablehead{
\colhead{Component} & \multicolumn{2}{c}{$I_{\mathrm{peak}}$ [Jy~beam$^{-1}$]} & \colhead{$\alpha_{43}^{86}$} \\
\cline{2-3} \\
\colhead{}          & \colhead{43.2~GHz} & \colhead{86.2~GHz}                  & \colhead{}}
\startdata
C1 & $6.28 \pm 0.02$ & $5.82 \pm 0.08$ & $-0.11 \pm 0.35$ \\
C3 & $6.82 \pm 0.02$ & $4.71 \pm 0.08$ & $-0.54 \pm 0.30$ \\
N1 & $0.18 \pm 0.02$ & $0.41 \pm 0.08$ & $+1.19 \pm 0.43$
\enddata
\tablecomments{Column 1: component name; Columns 2 and 3: peak intensity of a
restored image at 43.2~GHz with KaVA with the restored beam size of 1.36~mas
$\times$ 0.78~mas and the position angle of the major axis of 82\fdg3, and at
86.2~GHz with KVN; Column 4: spectral index.}
\end{deluxetable}
C1 and C3 show its spectral index of $-0.11 \pm 0.35$ and $-0.54 \pm 0.30$,
respectively, suggesting that C1 is the self-absorbed core and C3 is the
optically thin component,
whereas N1 has an optically thick spectral feature with $\alpha$ of $+1.19 \pm
0.43$ (see also Figure~\ref{fig:Figure2}).
Although optically thick spectral feature can also be seen in the northeast of
C1, we believe that this is not a real one because of lower signal-to-noise
ratio in the images at both frequencies at around this area compared to that
in the central region of the component.
To check fidelity of the spectral index map, we examined an effect of
registration error between images at each frequency taking into account the
expected amplitude error of the peak intensity of 15\% shown in
Section~\ref{sec:Section2}.
If we assume that the 15\% difference of the peak intensity comes from the
positional difference of the Gaussian component, possible registration
error of C3 becomes $\pm 0.32$~mas in the right ascension and $\pm 0.19$~mas
in the declination.
Assuming the maximum registration error between each image shown above,
resultant spectral index maps between 43 and 86 GHz show that $\alpha$ at the
peak position of N1 is inverted for all cases although an optically thick region
always appears at the edge of C1, probably due to lower signal-to-noise ratio.
We thus do not discuss a physical property at the edge of C1 in this paper.

The measured $\alpha$ for both C1 and C3 is consistent with previous results by
\citet{Suzuki12}, in which C1 is the radio core and C3 is the hot spot component
(i.e., a termination shock).
Comparison of $\alpha$ between C1 and N1 clearly indicates that N1 suffers
strong absorption from the intervening FFA foreground, which has been already
claimed by \citet{Fujita17}.

%%%%%%%%%%%%%%%%%%%%%%%%%%%%%
%%% Section 4: Discussion %%%
%%%%%%%%%%%%%%%%%%%%%%%%%%%%%
\section{Discussion}
\label{sec:Section4}

In this section, we discuss physical properties of the FFA foreground which
exists somewhere along our line of sight to N1.
The pioneering work of \citet{Walker00} showed that synchrotron emission from
the northern counter radio lobe located at the 5 -- 10~mas from C1 is obscured
by a FFA foreground for the first time.
In the present work, we investigate the newly emerged northern counter-jet
component N1 located around 2~mas from C1.
Using the observational properties of N1, here we investigate physical
properties of the FFA foreground.
While \citet{Walker00} discussed only a geometrically thin disk as the FFA
foreground, we argue both of a geometrically thin circumnuclear disk and assembly
of clumpy clouds as the FFA foreground in light of current understanding of more
realistic picture of AGN nucleus structure \citep{Wada12,Wada16,Izumi18}.
The considered structure is summarized in a schematic picture shown in
Figure~\ref{fig:Figure3}.
Hereafter we assume that the plasma composition of the FFA foreground is a pure
hydrogen.
Hence, an electron number density equals to that of protons.

%%%%%%%%%%%%%%%%%%%%%%%%%%%%%%%%%%%
%%% Subsection 4.1: FFA Opacity %%%
%%%%%%%%%%%%%%%%%%%%%%%%%%%%%%%%%%%
\subsection{FFA Opacity}
\label{subsec:Section4-1}

First, we estimate the FFA opacity for foreground seen at 86~GHz and 43~GHz.
It is well known that the theoretically known FFA opacity for the case of
uniformly distributed plasma is given by
\begin{equation}
\tau_{\rm ff}(\nu) \approx 25  \left(\frac{L}{1\,{\rm pc}}\right)
\left(\frac{n_{\rm e}}{10^{4}\,{\rm cm}^{-3}} \right)^{2}
\left(\frac{T_{\rm e}}{10^{4}\,{\rm K}} \right)^{-1.5}
\left(\frac{\nu}{1\,{\rm GHz}} \right)^{-2},
\label{eqn:Equation1}
\end{equation}
where $T_{\rm e}$ and $n_{\rm e}$ are the temperature and the electron density
of absorbing matter, respectively, and $L$ is the path length along the line of
sight \citep{Levinson95}.
Correspondingly, a change in the FFA depth, $\delta \tau_{\rm ff}$, for a given
frequency can be expressed as
\begin{equation}
\delta\tau_{\rm ff} = \left(\frac{\delta L}{L}
+ 2 \frac{\delta n_{\rm e}}{n_{\rm e}} - 1.5 \frac{\delta T_{\rm e}}{T_{\rm e}}
\right) \tau_{\rm ff}.
\label{eqn:Equation2}
\end{equation}
From this, one can discuss which quantity mainly contributes to the $\delta
\tau_{\rm ff}$ in different epochs.

Assuming that C3 and the corresponding counter-lobe component N1 have
intrinsically same intensities \citep{Fujita17}, one can obtain $\tau_{\rm ff}$
as
\begin{equation}
\exp \left[- \tau_{\rm ff}(\nu)\right] = \frac{I_{\rm N1}}{I_{\rm C3}},
\label{eqn:Equation3}
\end{equation}
where $I_{\rm C3}$ and  $I_{\rm N1}$ are the peak intensity of C3 and N1,
respectively.
Using Equation~(\ref{eqn:Equation3}) and the measured $I_{\rm N1}$ and
$I_{\rm C3}$, we obtain $\tau_{\rm ff} = 2.4 \pm 0.2$ at 86~GHz and $3.6 \pm 0.1$
at 43~GHz, respectively.
The ratio of $\tau_{\rm ff}$ between 43 and 86~GHz, which is given by
$\gamma = \log [{\tau_{\rm ff}({\rm 86GHz})} / \tau_{\rm ff}({\rm 43GHz})]/
\log{2} = -0.57 \pm 0.10$ is significantly different from $\gamma = -2$ which
indicates a non-uniformity of FFA even if the error of $\gamma$ is taken into
account.
Note that this well agrees to that derived between 15 and 43~GHz \citep{Fujita17}.
We also note that the spectral index map between the subsequent two frequencies
does not provide the turn-over frequency of FFA.
For its determination, VLBI observations at higher frequencies would be needed.

%%%%%%%%%%%%%%%%%%%%%%%%%%%%%%%%%%%%%%
%%% Subsection 4.2: FFA Foreground %%%
%%%%%%%%%%%%%%%%%%%%%%%%%%%%%%%%%%%%%%
\subsection{Physical Properties of FFA Foreground}
\label{subsec:Section4-2}

As mentioned in Section~\ref{subsec:Section4-1}, our observational results suggest
the existence of an optically thick, ionized non-uniform plasma foreground, which
may be an inner part of the rotational disk of the molecular gas within $\sim
100$~pc detected by ALMA \citep{Nagai19}, although our estimation of $\gamma$
cannot constrain an exact location of the FFA foreground.
As for the origin of FFA foreground, there are two possible cases to be realized.
One is (1) a CND in which the plasma density changes in the radial direction with
its size of 10~pc (see Figure~\ref{fig:Figure3}).
Another one is (2) an assembly of clumpy clouds, which are the main structure of
AGNs such as the broad-line region, narrow-line region, clumpy torus and polar
dust \citep[Figure~1 in][]{Ramos-Almeida17}.
Below we discuss physical properties of CNDs and clumps in
Sections~\ref{subsubsec:Section4-2-1} and \ref{subsubsec:Section4-2-2},
respectively.

%%%%%%%%%%%%%%%%%%%%%%%%%%%%%%%%%%%%%%%%%%%%%%%
%%% Subsubsection 4.2.1: Properties of CNDs %%%
%%%%%%%%%%%%%%%%%%%%%%%%%%%%%%%%%%%%%%%%%%%%%%%
\subsubsection{Properties of CNDs}
\label{subsubsec:Section4-2-1}

\begin{figure*}
\epsscale{1.00}
\plotone{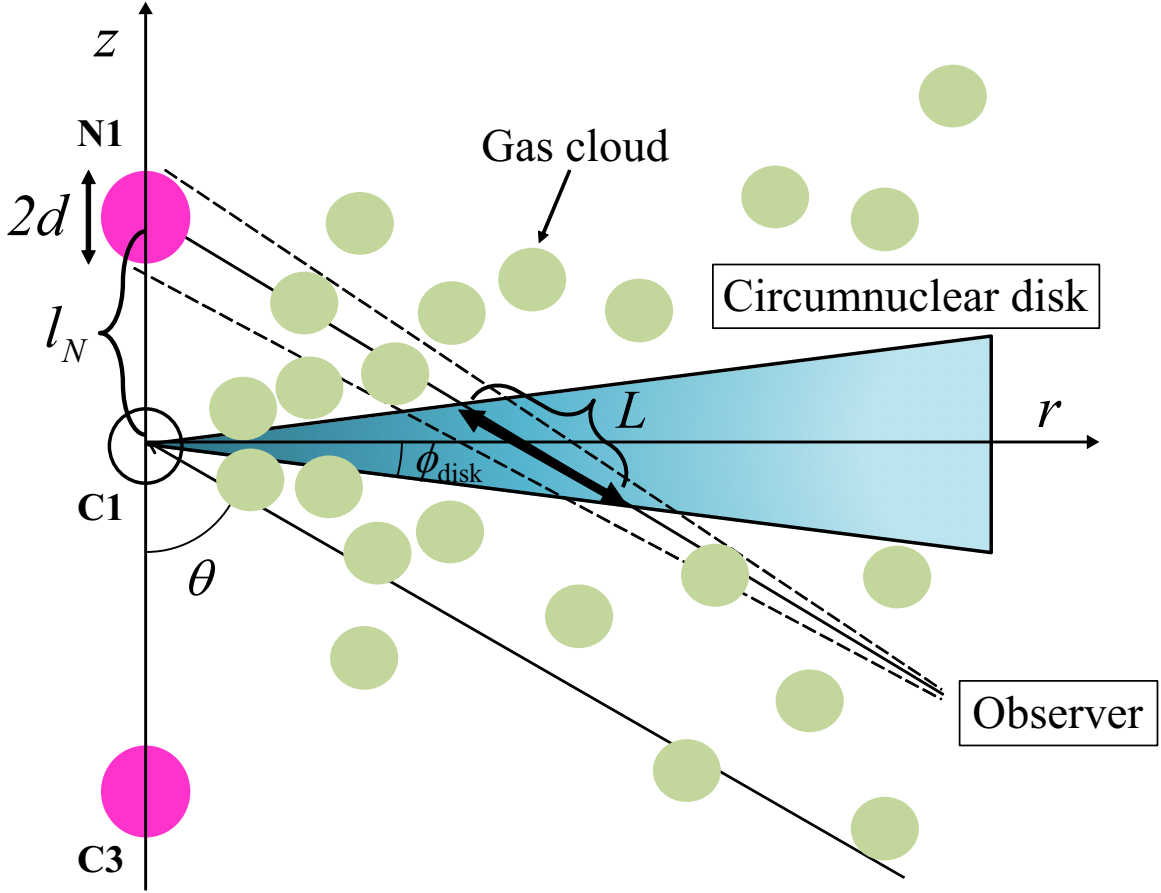}
\caption{A simple model of free-free absorption circumnuclear disk in which the
plasma density gradually changes in the radial direction.
The path length $L$ depends on $l_{\rm N}$ (apparent distance between C1 and N1),
$\theta$ (jet viewing angle), and $\phi_{\rm disk}$ (half-opening angle of the
disk).
The clumpy clouds with high velocity dispersion are located above the
circumnuclear disk.}
\label{fig:Figure3}
\end{figure*}

First, we discuss the case in which a circumnuclear disk with a constant
half-opening angle ($\phi_{\rm disk}$) is responsble for FFA (see
Figure~\ref{fig:Figure3}).
%%%%%%%%%%%%%%%%
%%% Figure 3 %%%
%%%%%%%%%%%%%%%%
The CND has a radial profile in $n_{\rm e}$.
The path length ($L$) and its change ($\delta L$) depend on the distance between
N1 and C1, $l_{\rm N}$ and the inclination angle of CND, $\theta$.
These are given by
\begin{equation}
L = \frac{\sin \theta \sin 2\phi}{\cos (\theta + \phi) + \cos (\theta - \phi)}
l_{\rm N},
\quad \frac{\delta L}{L} = \frac{\delta l_{\rm N}}{l_{\rm N}} \lesssim 0.4.
\label{eqn:Equation4}
\end{equation}
Therefore, the change of the geometric path length is not likely a main
contributor in $\delta \tau_{\rm ff}$ (see Equation~\ref{eqn:Equation2}) within
our observational period.
It is known that $l_{\rm N} = 2.2 (\sin \theta)^{-1}$~pc and the range of
$\theta$ is constrained as $18^{\circ} < \theta < 65^{\circ}$ by
\citet{Tavecchio14} and \citet{Fujita17}.
Here, the minimum half opening angle is assumed as $\phi_{\rm disk} = 0\fdg5$
(i.e., scale height $h \simeq 0.01r$, where $r$ is a radial distance from C1)
based on a hydrostatic ionized disk structure
\citep[see Section~3 in][]{Levinson95}.
Such a geometrically thin disk is observationally indicated from the large value
of the rotation measurement at 230~GHz \citep{Plambeck14}.
As for maximum $\phi_{\rm disk}$, we set the condition of $\phi_{\rm disk} <
90^{\circ} - \theta$, since the radio emission from C1 is not absorbed by the
foreground CND \citep[see Figure~\ref{fig:Figure2};][]{Plambeck14,Kim19}.
Then, we obtain the corresponding allowed range of $L$ in Table~\ref{tbl:Table2},
i.e., $0.04~{\rm pc} \lesssim L \lesssim 23~{\rm pc}$.
%%%%%%%%%%%%%%%
%%% Table 2 %%%
%%%%%%%%%%%%%%%
\begin{deluxetable}{cccc}
\tablecolumns{4}
\tabletypesize{}
\tablewidth{0pt}
\tablecaption{Path length $L$ for various models.\label{tbl:Table2}}
\tablehead{
\colhead{$\phi_{\rm disk}$ [$^{\circ}$]} & \multicolumn{3}{c}{$L$ [pc]} \\
\cline{2-4} \\
\colhead{} & \colhead{$\theta = 18^{\circ}$} & \colhead{$\theta = 40^{\circ}$} & \colhead{$\theta = 65^{\circ}$}}
\startdata
\phn\phn\phd 0.5 & \phd\phd 0.042     & \phd\phd 0.065     & \phd\phd 0.22 \\
\phn 5           & \phn\phd 0.43 \phd & \phn\phd 0.66 \phd & \phd\phd 2.23 \\
10               & \phn\phd 0.86 \phd & \phn\phd 1.35 \phd & \phd\phd 5.07 \\ 
15               & \phn\phd 1.31 \phd & \phn\phd 2.12 \phd & \phd\phd 9.85 \\ 
20               & \phn\phd 1.80 \phd & \phn\phd 3.01 \phd & \phd 22.9 \phd \\
25               & \phn\phd 2.32 \phd & \phn\phd 4.13 \phd & *         \\
30               & \phn\phd 2.91 \phd & \phn\phd 5.66 \phd & *         \\
35               & \phn\phd 3.59 \phd & \phn\phd 8.02 \phd & *         \\
40               & \phn\phd 4.41 \phd & 12.5 \phd & *         \\
45               & \phn\phd 5.44 \phd & *         & *         \\
50               & \phn\phd 6.82 \phd & *         & *         \\
60               & 12.3 \phd & *         & *
\enddata
\tablecomments{*\,: C1 is hidden by CNDs.}
\end{deluxetable}
The allowed path length is fairly wide due to the wide range of the allowed
$\theta$.
For instance, the lower limit of $L \approx 0.04$~pc realizes with the narrowest
viewing angle of $\theta = 18^{\circ}$.
This lower limit of $L \approx 0.04$~pc would correspond to the geometrical
thickness of the innermost part of CND.

\begin{figure*}
\epsscale{1.00}
\plotone{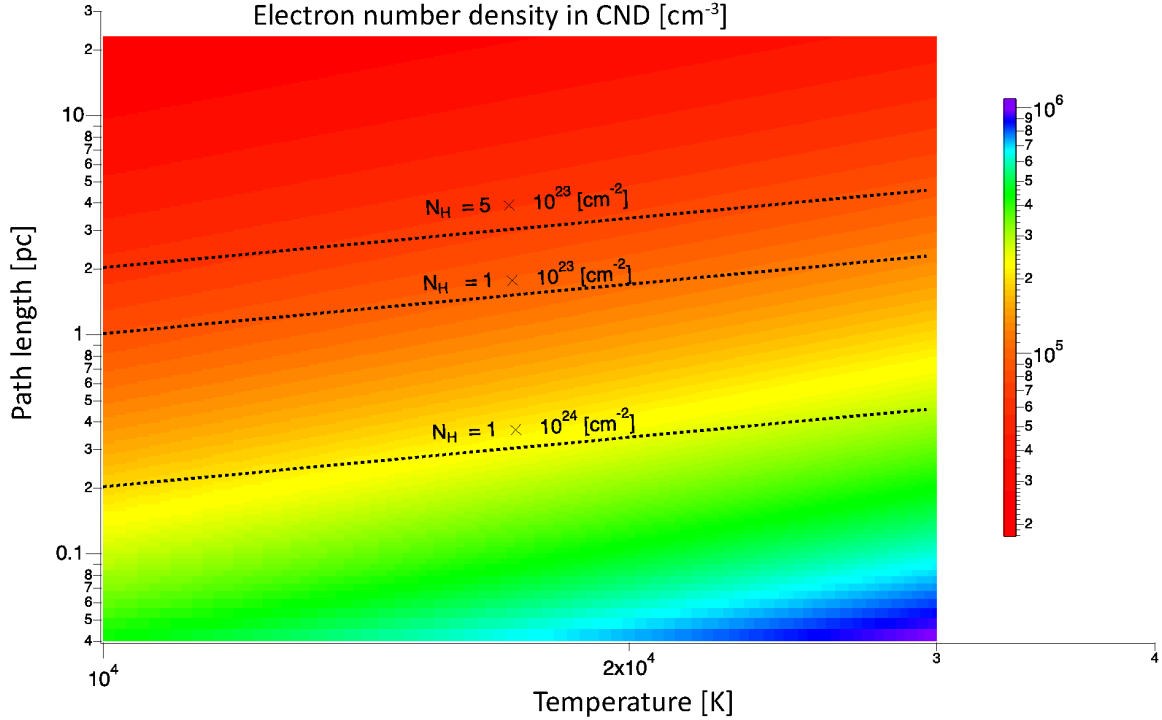}
\caption{Estimated electron number density ($n_{\rm CND}$) in case of the
circumnuclear disk (CND) being the FFA foreground.
The color bar represents the estimated $n_{\rm CND}$ for given $T_{\rm e}$ and
$L$, and the value is in the range of $1.8 \times 10^{4}~{\rm cm^{-3}} \lesssim
n_{\rm CND} \lesssim 1.0 \times 10^{6}~{\rm cm^{-3}}$.
The three dashed lines indicate the cases corresponding to the constant column
density, $N_{\rm H}$, of $1 \times 10^{23}~{\rm cm^{-2}}$, $5 \times
10^{23}~{\rm cm^{-2}}$, and $1 \times 10^{24}~{\rm cm^{-2}}$.}
\label{fig:Figure4}
\end{figure*}

In Figure~\ref{fig:Figure4}, we show the $n_{\rm e}$ of the CND, which satisfies
$\tau_{\rm ff}=1$ for given $L$ and $T_{\rm e}$.
%%%%%%%%%%%%%%%%
%%% Figure 4 %%%
%%%%%%%%%%%%%%%%
Hereafter we do not treat the vertical structure of the disk but just discuss an
averaged number density ($n_{\rm CND}$) along the given $L$.
The derived number density resides in
\begin{equation}
1.8 \times 10^{4} \, {\rm cm^{-3}} \lesssim n_{\rm CND} \lesssim
1.0 \times 10^{6} \, {\rm cm^{-3}},
\label{eqn:Equation5}
\end{equation}
properly taking possible uncertainties of the CND's $L$ and $T_{\rm e}$ into
account.
Regarding $T_{\rm e}$, the ionization condition requires the lower limit of
$T_{\rm e}$ as $T_{\rm e} \approx 1 \times 10^{4}$~K.
The upper limit of $T_{\rm e}$ is governed by atomic line cooling in the FFA
foreground \citep{Levinson95}.
Since the line cooling function is peaked around $T_{\rm e} \approx 1 \times
10^{4}$~K \citep{Sutherland93}, the assumed upper limit of $T_{\rm e}$ should not
be significantly different from $T_{\rm e} \approx 1 \times 10^{4}$~K.
Following \citet{Levinson95}, we set the upper limit as $T_{\rm e} \approx
3 \times 10^{4}$~K in this work.

The column density, $N_{\rm H}$, is more convenient quantity than $n_{\rm CND}$
for comparison of the obtained $n_{\rm CND}$ with numerical simulations and other
observations.
In Figure~\ref{fig:Figure4}, we overlay several lines with its $N_{\rm H}$ being
constant, as given by
\begin{equation}
N_{\rm H} \approx  3 \times 10^{23} \,{\rm cm}^{-2} \left(
\frac{n_{\rm CND}}{10^{5} \, {\rm cm}^{-3}} \right) \left( \frac{L}{1 \,{\rm pc}}
\right) .
\label{eqn:Equation6}
\end{equation}
\citet{Wada16} examined the structure and dynamics of molecular, atomic, and
ionized gases around an AGN by using three-dimensional radiation-hydrodynamic
simulations.
They found that inhomogeneous ionized gas are a geometrically thick, while dense
molecular gases are distributed near the equatorial plane. 
If the viewing angle for the nucleus is larger, e.g., $\theta \geq 50^{\circ}$,
the column density is consistent with our observation.
Thus, our observation might trace the parts of ionized gas demonstrated by
radiation-hydrodynamic simulations.
It is also interesting to note that \citet{Hitomi18} reported the detection of
the Fe-K$\alpha$ fluorescence line at 6.4~keV  from 3C~84 with the Soft X-ray
Spectrometer on board the Hitomi satellite with its equivalent width of
$\sim 20$~eV.
They derived $N_{\rm H}$ as $N_{\rm H} f_{\rm cov} \sim 3.0 \times
10^{22}$~cm$^{-2}$, where $f_{\rm cov}$ is the covering fraction of the
fluorescing material, from the Hitomi observation.
They discuss a possible matter distribution in case of the fluorescing material
being located at a distance of 100~pc from the central engine.
A small $f_{\rm cov}$ ($\sim 0.02$) can be derived if their result accommodates
to the electron density of the [\ion{Fe}{2}] emitters ($\sim 4000$~cm$^{-3}$)
obtained with the NIR observation by Gemini \citep{Scharwachter13}.
When adopting this $f_{\rm cov}\sim 0.02$, $N_{\rm H}$ becomes $\sim 1 \times
10^{24}$~cm$^{-2}$, which is comparable to our results with $L$ of a few pc and
$n_{\rm e}$ of a few $\times 10^5$~cm$^{-3}$ shown in Figure~\ref{fig:Figure4}.
It may indicate a possibility that the Hitomi satellite detected the fluorescence
line from a few pc region of CNDs that absorbed the synchrotron emission from N1.

It is important to verify that the picture described above is consistent with the
non-detection of N1 in the last decade.
Since $\delta L/L$ is too small to change $\delta \tau_{\rm ff}$ in our
observational period, $\delta n_{\rm CND}/n_{\rm CND}$ would be a main
contributor of $\delta \tau_{\rm ff}$.
It is expected that $n_{\rm CND}$ is larger in earlier observational epochs.
As an example, we reanalyzed one epoch of VLBA archive data at 43~GHz conducted 
on 2012 October 29, which showed non-detection of N1 \citep{Jorstad17}.
The data shows the peak intensity of the southern component of 1.94~Jy~beam$^{-1}$
with the 1-$\sigma$ image noise of 4.2~mJy~beam$^{-1}$.
If we set the detection criterion of N1 as 5 times the image noise, the upper
limit of the peak intensity of N1 shall become 21~mJy~beam$^{-1}$, resulting in
the lower limit of the opacity to be 4.5.
Assuming the typical values of $T_{\rm e}$ of $1 \times 10^4$~K and $L = 0.5$~pc
for 2012 October, $n_{\rm CND} \gtrsim 2.6 \times 10^5$~cm$^{-3}$ is required to
accommodate the opacity obtained by VLBA to the result of non-detection of N1.
On the other hand, taking the same typical values of $L \approx  0.5$~pc and
$T_{\rm e} \approx 1 \times 10^4$~K and applying $\tau_{\rm ff} = 3.6$ which was
taken by our KaVA observation in 2016, we obtain $n_{\rm CND} \approx 2.3 \times
10^5$~cm$^{-3}$, which is smaller than $n_{\rm CND}$ in 2012.
Although the change is too subtle to draw a conclusion, the comparison of
two-epoch images obtained in 2012 and 2016 indicates a slightly larger
$n_{\rm CND}$ for earlier epoch, which is consistent with a radial gradient in
the disk. 
To make a clear conclusion, further yearly timescale long term monitoring of 3C~84
is much awaited.

%%%%%%%%%%%%%%%%%%%%%%%%%%%%%%%%%%%%%%%%%%%%%%%%%%%%%%%%
%%% Subsubsection 4.2.2: Properties of Clumpy Clouds %%%
%%%%%%%%%%%%%%%%%%%%%%%%%%%%%%%%%%%%%%%%%%%%%%%%%%%%%%%%
\subsubsection{Properties of Clumpy Gas Clouds}
\label{subsubsec:Section4-2-2}

Second, we discuss the case in which clumpy gas clouds are responsible for FFA
(the clouds in Figure~\ref{fig:Figure3}).
This idea is motivated by recent observations of 10~pc obscuring structures in
nearby AGNs with ALMA \citep{Garcia-Burillo14,Imanishi16,Imanishi18,Izumi18}.
The detailed obscuring structures revealed by ALMA consist of not only a dusty
disk, but also clumpy gas clouds with high velocity dispersion implied by
supernovae and/or AGN radiative feedbacks
\citep[e.g.,][]{Wada16,Izumi18,Kawakatu20}.
Hence, we consider the case in which clumpy gas clouds dominantly absorb the
synchrotron radio emission from N1.
In the clouds-dominated case, the FFA optical depth can be written as
\begin{eqnarray}
\tau_{\rm ff} & = & \bar{\tau}_{\rm ff} N_{\rm c} \nonumber \\
& \approx & 25  \left( \frac{N_{\rm c}}{10} \right)
\left( \frac{r_{\rm c}}{0.1 \, {\rm pc}} \right)
\left( \frac{\bar{n}_{\rm c}}{10^{4} \, {\rm cm}^{-3}} \right)^{2}
\left( \frac{T_{\rm e}}{10^{4} \, {\rm K}} \right)^{-1.5}
\left( \frac{\nu}{1 \, {\rm GHz}} \right)^{-2},
\label{eqn:Equation7}
\end{eqnarray}
where $\bar{\tau_{\rm ff}}$, $N_{\rm c}$, $r_{\rm c}$ and $\bar{n}_{\rm c}$ are
the average FFA optical depth of each cloud, the total number of clouds in the
line of sight and the size of a cloud, and the number density of each cloud,
respectively.
Here, we note that $\bar{\tau} = \alpha_{\rm ff} r_{\rm c}$ where
$\alpha_{\rm ff}$ is the free-free absorption coefficient.

Assuming $N_{\rm c}$ is comparable to the mean number of clouds along radial
equatorial ray for Seyfert galaxies derived by the IR SED fitting
\citep[e.g.,][]{Alonso-Herrero11,Ramos-Almeida11,Ichikawa15,Audibert17},
i.e., $N_{\rm c}\simeq 3 - 15$, we find that $\bar{\tau}_{\rm ff}$ at 86~GHz
becomes $\simeq 0.2 - 1.2$, which is of the order of unity and agrees with
observational properties of N1.

To constrain the number density of the clump, getting the size of the clump
($r_{\rm c}$) should be required.
The size of self-gravitating clump in the context of AGNs have been estimated in
literatures \citep[e.g.,][]{Krolik88,Honig07,Kawaguchi11}.
According to them, we have $r_{\rm c} \leq 0.05$~pc at 1~pc from the
central SMBH with $M_{\rm BH} = 8 \times 10^{8} M_{\odot}$ \citep{Scharwachter13}.
On the other hand, the size of clumps can be constrained by the observations of
transient X-ray absorption events in nearby AGNs, i.e., the typical size is
0.002~pc \citep[e.g.,][]{Markowitz14,Tanimoto19}.
By adopting $N_{\rm c}$ and $r_{\rm c}$ taking into account their uncertainties
of $3 \leq N_{\rm c} \leq 15$ and $0.02 \, {\rm pc} \leq r_{\rm c} \leq
0.05 \, {\rm pc}$, the lower limit of number density of each ionized cloud
$\bar{n}_{\rm c}$ can be given by
\begin{equation}
3\times 10^{5}~{\rm cm}^{-3} \leq
\bar{n}_{\rm c} \leq 4\times 10^{6}~{\rm cm}^{-3}.
\label{eqn:Equation8}
\end{equation}
In case of ionized gas clumps, the typical Thomson scattering opacity of each
cloud can be estimated as $\bar{\tau}_{\rm T}\sim 1.5\times 10^{-2}$ since the
optical depth is given by $\tau_{\rm T} = \sigma_{\rm T} \bar{n_{\rm c}}
r_{\rm c}$, where $\sigma_{\rm T}$ is the Thomson cross section.
If we adopt the opacity ratio of the dusty gas and ionized gas is $\simeq 10^{3}$
at the UV band for the AGN radiation \citep[e.g.,][]{Umemura98,Ohsuga01,Wada12},
the optical depth of each cloud is $\bar{\tau}_{\rm UV} = 10^{3} \times
\bar{\tau}_{\rm T} \simeq 15$, which is consistent with the lower value for nearby
Seyfert galaxies \citep[e.g., Table~10 in][]{Ramos-Almeida11}.
This might indicate that the ionized gas is also clumpy within 1~pc region of
3C~84.
\citet{Wada18} examine properties of the ionized gas irradiated by less
luminous AGN such as Seyfert galaxies based on their ``radiation-driven
fountain'' model \citep{Wada12}.
They found that the ionized region show non-uniform internal structures,
corresponding to the clumpy fountain flows caused by the radiation pressure on
dusty gas, although the typical density ($\bar{n}_{\rm c} \simeq
10^{3}~{\rm cm}^{-3}$) is smaller than our estimate.
In addition,  by the optical/NIR observations, the existence of dense clumps with
$\bar{n}_{\rm c} \simeq 10^{5}$~cm$^{-3}$ has been reported from the detection of
coronal lines within NLR \citep[e.g.,][]{Murayama98}.
These high density clouds in NLRs might contribute the absorption feature of N1. 

Alternatively, since the density is comparable to number density $n_{\rm c} \simeq
10^{3-5}$~cm$^{-3}$, based on the momentum balance between the jet thrust and the
ram pressure from the clump \citep{Nagai17,Kino18}, these ionized clumps may
contribute not only the absorption of counter jet but also the feedback on jets.
Since the dust sublimation radius of 3C~84 is $\sim 0.1\,{\rm pc}\,(L_{\rm UV}/
10^{43} \, {\rm erg} \, {\rm s}^{-1})^{0.5}$ \citep[e.g.,][]{Kino18}, the dust
components in these dense clumps could be survived.
If this is the case, the clumpy clouds might be related to the polar elongation
in MIR continuum emission distributions revealed by high-resolution observations
in nearby Seyfert galaxies
\citep[e.g.,][]{Tristram14,Asmus16,Lopez-Gonzaga16}.

Lastly, it is worth mentioning that the size of clumpy clouds may be constrained
by the multi-epoch observation of FFA, which is left in our future work, since
the typical timescale of flux variability may be related to the hot spots
crossing time, i.e., $t_{\rm cross} \sim r_{\rm c} / v_{\rm h} \simeq 2.5$~yr
by assuming $r_{\rm c} = 0.05\,{\rm pc}$ and the head speed of jets $v_{\rm h}
\simeq 0.2 c$ \citep[e.g.,][]{Nagai10,Suzuki12,Hiura18}.
If we detect the flux variability with a few years, it may clarify whether the
clumpy clouds are main absorbers rather than the CNDs.

%%%%%%%%%%%%%%%%%%%%%%%%%%
%%% Section 5: Summary %%%
%%%%%%%%%%%%%%%%%%%%%%%%%%
\section{Summary}
\label{sec:Section5}

By conducting quasi-simultaneous VLBI observations at 43 and 86~GHz with KaVA
and KVN, we explore sub-parsec scale structure of a nearby bright radio galaxy
3C~84 via the optically thick FFA features.
Here we summarize our main findings.

\begin{itemize}

\item
We conducted a new quasi-simultaneous observation of 3C~84 with KVN at 86~GHz and
KaVA at 43~GHz in 2016 February.
We succeeded the first detection of N1 at 86~GHz and the data show that N1 still
has an inverted spectrum between 43 and 86~GHz with its spectral index $\alpha$
($S_{\nu} \propto \nu^{\alpha}$) of $1.19 \pm 0.43$, while the approaching lobe
component C3 has the steep spectrum with $\alpha$ of $-0.54 \pm 0.30$.

\item
The opacity of FFA is less dependent on frequency than the case for uniform
absorbers, i.e., $\tau_{\rm ff} \propto \nu^{-0.57 \pm 0.10}$.
Thus, it suggests that a absorbing medium would be a highly inhomogeneous
structure and it is consistent with the previous work of \citet{Fujita17}.

\item
Based on the measured flux asymmetry between the counter and approaching lobes,
we constrain the number density of the FFA foreground $n_{\rm CND}$ as $1.8
\times 10^{4}~{\rm cm^{-3}} \lesssim n_{\rm CND} \lesssim 1.0 \times
10^{6}~{\rm cm^{-3}}$, considering the uncertainties of temperature and path
length of CNDs having gradual change in the plasma density.
We also discuss the case of non-uniform CNDs containing clumpy clouds.
By considering the size of clouds with sub-pc, we constrain the number density
of cloud with $3.0 \times 10^{5}~{\rm cm}^{-3} \leq \bar{n}_{\rm c} \leq
4.0 \times 10^{6}~{\rm cm}^{-3}$.
In both cases, the derived electron number density is higher than the typical
value ($\bar{n}_{\rm c} = 10^{2-4}\,{\rm cm}^{-3}$) seen in narrow-line region,
suggesting that such dense ionized clumps might be located at pc-scale central
region of 3C~84.

\end{itemize}

%%%%%%%%%%%%%%%%%%%%%%%
%%% Acknowledgments %%%
%%%%%%%%%%%%%%%%%%%%%%%
\acknowledgments

We are grateful to the anonymous referee for valuable comments, which improved
the manuscript.
We acknowledge all staff members and students at KVN and VERA who supported the
operation of the array and the correlation of the data.
KVN is a facility operated by the Korea Astronomy and Space Science Institute.
VERA is a facility operated by National Astronomical Observatory of Japan in
collaboration with associated universities in Japan.
This work is partially supported by JSPS KAKENHI Grant Numbers JP18K03656 and
JP18H03721 (MK).
NK acknowledges the financial support of Grant-in-Aid for Young Scientists
(B:16K17670) and Grant-in-Aid for Scientific Research (C:19K03918).

%%%%%%%%%%%%%%%%%%
%%% References %%%
%%%%%%%%%%%%%%%%%%

\end{document}